\documentclass[12pt,preprint]{aastex}
\usepackage{amsmath}
\usepackage{subfigure}
\usepackage{rotating}
\newif\ifAMStwofonts
\AMStwofontstrue

\def\arcdeg{\hbox{$^\circ$}}

\def\ga{\mathrel{\hbox{\rlap{\hbox{\lower4pt\hbox{$\sim$}}}\hbox{$>$}}}}
\def\la{\mathrel{\hbox{\rlap{\hbox{\lower4pt\hbox{$\sim$}}}\hbox{$<$}}}}

\shorttitle{Relativistic Binary Pulsar PSR B1913+16}

\shortauthors{J. M. Weisberg  et al.}

\begin{document}

\title{Timing Measurements of the Relativistic Binary Pulsar \\
  PSR B1913+16}

\author{J. M. Weisberg}
\affil{Department of Physics and Astronomy, Carleton College, Northfield, MN 55057}
\email{jweisber@carleton.edu}
\author{D.J. Nice}
\affil{Department of Physics, Lafayette College, Easton, PA 18042}
\email{niced@lafayette.edu}
\and
\author{J. H.  Taylor}
\affil{Joseph Henry Laboratories and Department of Physics, 
  Princeton University, Princeton, NJ 08544}
\email{joe@princeton.edu}

\begin{abstract}

We present results of more than three decades of timing measurements
of the first known binary pulsar, PSR B1913+16.  Like most other
pulsars, its rotational behavior over such long time scales is
significantly affected by small-scale irregularities not explicitly
accounted for in a deterministic model.  Nevertheless, the physically
important astrometric, spin, and orbital parameters are well
determined and well decoupled from the timing noise.  We have
determined a significant result for proper motion, $\mu_{\alpha} =
-1.43\pm0.13$, $\mu_{\delta}=-0.70\pm0.13$ mas yr$^{-1}$.  The pulsar
exhibited a small timing glitch in May 2003, with ${\Delta
  f}/f=3.7\times10^{-11}$, and a smaller timing peculiarity in
mid-1992.  A relativistic solution for orbital parameters yields
improved mass estimates for the pulsar and its companion,
$m_1=1.4398\pm0.0002 \ M_{\sun}$ and $m_2=1.3886\pm0.0002
\ M_{\sun}$.  The system's orbital period has been decreasing at a
rate $0.997\pm0.002$ times that predicted as a result of gravitational
radiation damping in general relativity.  As we have shown before,
this result provides conclusive evidence for the existence of
gravitational radiation as predicted by Einstein's theory.
\end{abstract}

\slugcomment{Accepted to APJ, published in vol. 722, p. 1030 (2010)}


\setcounter{footnote}{0}

\section{Introduction}

Pulsar PSR B1913+16 (PSR J1915+1606) was the first binary pulsar to be
discovered \citep{ht75}.  Its 7.75-hr binary period and 300
km~s$^{-1}$ projected orbital velocity suggested that it would exhibit
a rich set of potentially measurable relativistic effects, and that
has turned out to be the case.  Subsequent studies have shown that the
pulsar's companion must also be a neutron star.  With tidal and
rotational stellar distortions eliminated as complicating factors for
any dynamical analysis, the system is an essentially clean
laboratory for testing relativistic gravity.  In this respect, the
largest remaining complications depend on reference-frame
accelerations related to the structure and dynamics of our Galaxy.

Over the past 35 years, we and our colleagues have timed PSR~B1913+16
at Arecibo Observatory with steadily improving equipment and analysis
techniques.  Among the best known results are measurement of the
general relativistic advance of periastron at a rate some 35,000 times
that of Mercury in the solar system \citep{tet76}; the effect of
gravitational radiation damping, causing a measurable rate of orbital
decay \citep{tet79}; and detection of changes in the pulse shape,
resulting from geodetic spin precession \citep{wet89}.  Our results
continue to be fully consistent with general relativity and have
placed strong constraints on alternative, previously viable,
relativistic theories of gravity
\citep{wt81,tw82,tw89,tet92,wt02,cw08}.  The purpose of this paper is
to provide our latest analysis of the pulsar and its orbit, based on
timing observations from 1974 through 2006.

\section{Observations}

Our highest quality dataset consists of 7650 five-minute integrations
of the pulse profile at wavelengths near 20~cm, each subsequently
reduced to a topocentric time-of-arrival (TOA) measurement at Arecibo
Observatory.  These data span the years 1981 through 2006; earlier
measurements are consistent with these but (owing to their much larger
uncertainties) have little effect on the final results.  In 2004 we
made a special effort to observe regularly throughout the year, so as
to improve quality of the astrometric results.  Parameters of the
observing equipment and associated measurement uncertainties are
presented in Table \ref{table:obsys}.  Further details of the Princeton Mark I
observing system are given by \citet{tw82}, while the Mark II and III
are described in \citet{tw89}.  The latest system, the Wideband
Arecibo Pulsar Processor (WAPP), consists of four spectrometers
optimized for pulsar work, each having 512 frequency channels across a 100
MHz bandwidth and synchronously accumulating the pulsar signal into
1024 phase bins across the full pulsar period.

\section{Analysis of TOAs}

Data were analyzed with the program TEMPO, using the JPL DE405 solar
system ephemeris and the relativistic timing model of \citet{dd86}.
The Damour-Deruelle approach is particularly desirable because it
leaves open the question of the correct relativistic theory of
gravity: one solves for phenomenological parameters whose precise
relation to a gravitational theory can be investigated later.  Further
discussion of the Damour-Deruelle model and its application to binary
pulsars in TEMPO can be found in \citet{tw89}.  Important quantities
determined in this way include those related to the pulsar's celestial
position, spin, and orbit.  The first category includes right
ascension $\alpha$, declination $\delta$, proper motions
$\mu_{\alpha}$ and $\mu_{\delta}$, epoch $t_0$, pulse repetition
frequency $f$, spindown rate $\dot{f}$, and a glitch epoch and frequency 
discontinuity $\Delta f$.  Fitted orbital
parameters include five Keplerian quantities: the projected semimajor
axis  of the pulsar orbit $x\equiv a_1 \sin i$, eccentricity $e$, epoch of periastron
passage $T_0$, period $P_b$, and longitude of periastron $\omega_0$;
and the following relativistic or ``post-Keplerian'' parameters:
average rate of periastron advance $\langle\dot\omega\rangle$,
variations in gravitational redshift and time dilation $\gamma$, and
orbital period derivative $\dot{P_b}$.  As with most other pulsars
that have been timed carefully over several decades, a number of
``nuisance parameters'' must also be measured to account for unmodeled
long-term timing irregularities.  These extra fitted terms have no
clear physical interpretation beyond being somehow related to the
poorly understood structure and dynamics of the spinning neutron star.  Their
mathematical form is somewhat arbitrary; as described further below,
we have experimented with a
number of higher-order time derivatives of the pulsar rotation
frequency, and one additional frequency discontinuity.

\subsection{Astrometric and Spin Parameters}
\label{sec:astroresults}

The astrometric and basic spin parameters determined from the full data set are
listed in Table~\ref{table:astromfits}.  A significant result for
proper motion has been obtained for the first time.  While the pulsar position is determined 
from the annual variation  of  TOAs  as the Earth moves about its orbit,  proper motion
is determined from  a secular variation of this annual signal. Timing noise or other systematic
effects  can contaminate such measurements. We now believe that the previously reported value 
of PSR B1913+16's proper motion \citep{tw89} was biased by timing noise, a problem which 
was exacerbated by the concentrated  but $\sim$biennially spaced  observing campaigns 
frequently employed for this source.  To circumvent these problems, we observed 
PSR B1913+16 several times over the course of 
calendar year 2004 to achieve thorough data coverage around the Earth's orbit. By merging the
2004 data with observations made in 1985-1988, which also had good coverage throughout
those years, 
we have now  obtained a robust measurement of the pulsar's proper motion.  See Section
\ref{sec:vel} for further analyses of the proper motion result.

As in other pulsars, the measured spindown rate $\dot f$ is assumed to
be the result of an electromagnetic braking torque on the spinning,
strongly magnetized neutron star.  Its value is deterministic and
(at the level of accuracy quoted in Table~2) 
independent of the span of data over which it is fitted.
 In addition,the pulsar experienced a well-defined ``classical'' glitch in May 2003. 
The magnitude $\Delta
f/f=3.7\times10^{-11}$ of the event is smaller than almost all seen in the population
of normal (not recycled) pulsars, with only some of the glitches in PSR B0355+54
having a comparably small magnitude
 \citep{mel08,cu10}. Ours is only the second glitch to be detected in a recycled pulsar, 
with the other (the smallest known among all pulsars) occurring in the millisecond 
pulsar PSR B1821$-$24 in globular cluster M28 \citep{cb04}. 

The remaining timing parameters --- higher-order derivatives $\ddot
f$, $\dddot f$, \ldots,\ and (in some fits) another small frequency discontinuity
 in mid-1992 --- were introduced
to the timing solution in an {\it ad hoc} manner, in order to
``whiten'' the remaining post-fit residuals.   Although we offer no 
clear or unique physical
interpretation for these parameters, their combined effects
are almost certainly a consequence of stochastic
timing-noise processes in the neutron star interior
\citep{cd85,aet94,ulw06}. Unlike the case of the well-fitted May 2003
glitch, the
values of these fitted parameters are  not independent of the data span analyzed
and cannot be expected to extrapolate the timing behavior
accurately to future epochs.  Identifying the mid-1992 behavior
 as a discrete event is highly uncertain, in part because of 
coarse sampling around that time.  Our data can be fit almost as well
by introducing several additional frequency derivatives instead of a
second discrete event.   Similar results were obtained by fitting multiple harmonically
related sinusoids to the timing noise with the routine ``FITWAVES'' \citep{het04} 
of program TEMPO2 \citep{het06,eet06}, instead of the multiple frequency 
derivatives described above. Since we are unable to converge on a unique glitch parameter 
solution for the mid-1992 behavior, we  do not include  this event
in Table~\ref{table:astromfits}. 

\subsection{Orbital Parameters}
\label{sec:orbresults}

Fitted orbital parameters for PSR B1913+16 are listed in Table
\ref{table:orbfits}.  As we have emphasized before \citep{tw89},
values for each of the Damour-Deruelle post-Keplerian parameters 
expected in general relativity can be expressed in terms of the 
Keplerian parameters and
the initially unknown masses of the pulsar and its companion, $m_1$
and $m_2$.  The appropriate expressions for $\langle\dot\omega\rangle$
and $\gamma$ are
\begin{eqnarray}
\langle\dot{\omega}\rangle
    &=& 3 \  G^{2/3} \  c^{-2} \
(P_b / 2 \pi)^{-5/3} \  (1-e^2)^{-1} \ (m_1 + m_2)^{2/3} \nonumber \\
    &=& 2.113323(2)\ \left[\frac{(m_1 + m_2)}{M_\sun}\right]^{2/3}
    {\rm deg~yr}^{-1}, \\
\gamma
   &=& G^{2/3} \  c^{-2} \ e \
(P_b / 2 \pi)^{1/3} \ m_2 \ (m_1 + 2 m_2) \ (m_1 + m_2)^{-4/3} \nonumber  \\
   &=& 0.002936679(2) \  \left[\frac{m_2 \ (m_1 + 2 m_2) \ (m_1 + m_2)^{-4/3}}{M_\sun^{2/3}}\right]~{\rm s}.
\end{eqnarray}
In the second line of each equation we have substituted values for
$P_b$ and $e$ from Table \ref{table:orbfits}, and used the constants $G
M_{\sun}/c^3=4.925490947\times10^{-6}$~s and 1 Julian yr $=86400
\times 365.25$~s.  The figures in parentheses represent uncertainties
in the last quoted digit, determined by propagating the uncertainties listed
in Table \ref{table:orbfits}. In each case, the uncertainties are  dominated by the experimental
uncertainty in orbital eccentricity, $e$.

Eq.~(1) may be solved for the total mass of the PSR B1913+16 system,
yielding $M=m_1 + m_2=2.828378 \pm 0.000007~M_{\sun}$.  The additional
constraint privided by Eq.~(2) permits a solution for each star's mass
individually, $m_1=1.4398\pm0.0002 \ M_{\sun}$ and
$m_2=1.3886\pm0.0002 \ M_{\sun}$.  As far as we know, these are the
most accurately determined stellar masses outside the solar system.
It is interesting to note that since the value of Newton's constant
$G$ is known to a fractional accuracy of only $1\times10^{-4}$, $M$
can be expressed more accurately in solar masses than in grams.

\subsection{Gravitational Radiation Damping}

\label{sec:radresults}

According to general relativity a binary star system should radiate
energy in the form of gravitational waves.  Peters and Matthews (1963)
showed that the resulting rate of change in orbital period should be
\begin{eqnarray}
\dot{P}_b^{\rm GR} &=&-\frac{192\,\pi\,G^{5/3} } {5\,c^5}
\left(\frac{P_b}{2\pi}\right)^{-5/3} 
  \left(1 + \frac{73}{24} e^2 + \frac{37}{96} e^4\right) (1-e^2)^{-7/2} 
\ m_1\,m_2\,(m_1+m_2)^{-1/3} \nonumber \\
&=& -1.699451(8)\times 10^{-12} \left[\frac{m_1\, m_2\, (m_1+m_2)^{-1/3}}
  {M_\sun^{5/3}}\right] 
\end{eqnarray}
Inserting values obtained for $m_1$ and $m_2$ and propagating
uncertainties appropriately, we obtain the general relativistic
predicted value
\begin{equation}
\dot{P}_b^{\rm GR} = -2.402531\pm0.000014 \times 10^{-12}.
\end{equation}

\begin{figure}
\includegraphics[trim=0in 0in  0 0 ,clip,scale=0.9]{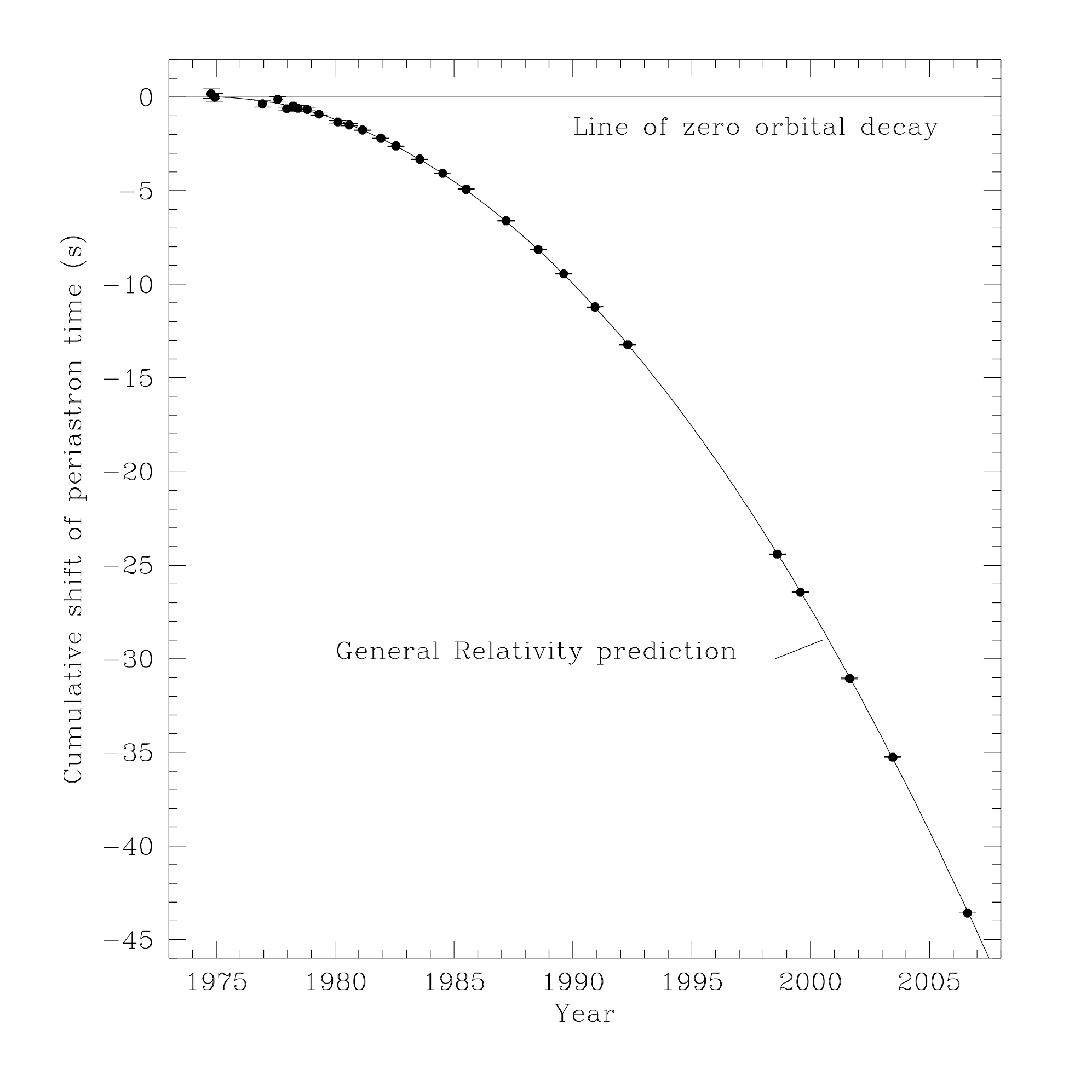}

\caption{Orbital decay caused by the loss of energy by gravitational
  radiation.  The parabola depicts the expected shift of periastron
  time relative to an unchanging orbit, according to general
  relativity.  Data points represent our measurements, with error bars
  mostly too small to see.}

\label{fig:parabola}
\end{figure}

Equations (3) and (4) apply in the orbiting system's reference frame.
Relative acceleration of that frame with respect to the solar system
barycenter will cause a small additional contribution to the observed
$\dot{P}_b$.  \citet{dt91} presented a detailed discussion of this
effect and other possible contributions to $\dot{P}_b$.  Recent
progress in determining the galactic-structure parameters allows us to
update the relevant quantities and compute a new value for the
kinematic correction to $\dot{P}_b$.  Using $R_0=8.4\pm 0.6$~kpc for
the distance to the galactic center and
$\Theta_0=254\pm16$~km~s$^{-1}$ for the circular velocity of the local
standard of rest \citep{ghez08,get09,ret09}, and $d=9.9\pm3.1$~kpc for
the pulsar distance \citep{wet08}, we obtain the kinematic
contribution, $\Delta\dot{P}_{\rm b, gal}$:
\begin{equation}
\Delta\dot{P}_{\rm b, gal}=-0.027\pm0.005 \times 10^{-12}\ .
\end{equation}
Thus, we find the ratio of observed to predicted rate of orbital
period decay to be
\begin{equation}
\frac{\dot{P}_b - \Delta\dot{P}_{\rm b, gal}}{\dot{P}_b^{\rm GR}} = 0.997\pm0.002.
\end{equation}
Agreement between the observed orbital decay and the general
relativistic prediction is illustrated in Fig.~\ref{fig:parabola},
which shows how excess orbital phase (relative to an unchanging orbit)
has accumulated since the pulsar's discovery in 1974.  We note that
the overall experimental uncertainty embodied in Eq.~(6) is now
dominated by uncertainties in the galactic parameters and pulsar
distance, not the pulsar timing measurements.  Even better agreement
between observed and expected values of $\dot{P}_b$ would be obtained
if the true value of $R_0$ or $d$ were slightly smaller, or $\Theta_0$
slightly larger. For example, observed and expected values agree if
$d=6.9$ kpc, which is within the \citet{wet08} error envelope.
 It will be interesting to see whether improved
future estimates of these quantities will show one or more of
these conditions to be true.

\section{Other Relativistic Effects}
Two other relativistic phenomena are potentially measurable in the PSR
B1913+16 system: geodetic precession and gravitational propagation
delay.  Spin-orbit coupling should cause the pulsar's spin axis to
precess \citep{dr74,boa75,bob75}, which should lead to observable
pulse shape changes.  \citet{wet89} first detected such changes, which
were observed and modeled further by \citet{k98}.  \citet{wt02} and
\citet{cw08} found that the pulsar beam is elongated in the latitude
direction and becomes wider in longitude with increasing distance from
the beam axis in latitude.  These models suggest that in the next
decade or so, precession may move the beam far enough that the pulsar
will become unobservable from Earth for some decades, before
eventually returning to view.

\label{sec:propresults}
The excess propagation delay \citep{sha64} caused by passage of pulsar
signals through the curved spacetime of the companion is largest at
the pulsar's superior conjunction.  The maximum amplitude varies with
time because the impact parameter at superior conjunction depends
strongly on the current value of $\omega$.  In this respect the 
orbital geometry was
particularly unfavorable in the mid-1990s (see Damour \& Taylor 1992),
but in coming years the propagation delay should start to become
observable.  \citet{dd86} characterize the measurable quantities as
range $r=(G m_2/c^3)$ and shape $s\equiv\sin i$, of the Shapiro delay,
where $i$ is the orbital inclination.  As orbital precession carries
our line of sight deeper into the companion's gravitational well,
future observations should permit the robust measurement of these two
parameters, and hence two additional tests of relativistic theories of
gravity \citep{d07,esp09}.

\section{Systemic Velocity}
\label{sec:vel}

Our pulsar  proper motion measurement (Section \ref{sec:astroresults}), combined with 
the distance estimate discussed in Section \ref{sec:radresults}, 
 corresponds to a transverse velocity  (with respect to the solar system barycenter) of 
 75~km$\,$s$^{-1}$ with a galactic position angle of $306\degr$;  i.e., directed 
 36 degrees above the galactic plane.  The $\sim30\%$ distance uncertainty places similar
 limits on  velocity accuracies.

We can now estimate two components of the pulsar systemic velocity  in {\it{its own}} standard of 
rest by combining the measured pulsar transverse velocity and distance,
the  solar motion with respect to our Local Standard of Rest \citep{set10} , and 
galactic quantities $R_0$ and $\Theta_0$.   
The third component of motion, which is inaccessible via proper motion measurements, lies 
close to the direction of Galactic rotation at the pulsar's position.

The pulsar's galactic planar and polar velocity components relative to its standard of
rest  are 247 km/s almost directly away from the galactic center and 51 km/s toward the 
galactic North Pole, respectively.  (This is significantly larger than the measured velocity in the 
solar system barycenter frame because  the pulsar's standard of rest velocity  fortuitously cancels 
much of the pulsar's peculiar velocity with respect to it.) The 
systemic velocity  of B1913+16 is significantly larger
 than other well-measured double neutron star binary system velocities, including the 
 J0737-3037 (transverse velocity 10 km/s; Stairs et al. 2006), J1518+4905 (transverse
 velocity 25 km/s; Janssen et al. 2008), and B1534+12 (transverse 
 velocity 122 km/s; Thorsett et al. 2005)  systems.

\section{Conclusions}

We have analyzed the full set of Arecibo timing data on pulsar
B1913+16 to derive the best values of all measurable quantities.  A
significant proper motion has finally been determined.  A small glitch
was observed in the pulsar's timing behavior, the second known glitch
in the population of recycled pulsars.  The measured rate of orbital
period decay continues to be almost precisely the value predicted by
general relativity, providing conclusive evidence for the existence of
gravitational radiation.  Uncertainties in galactic accelerations now
dominate the error budget in $\dot P_b$, and are likely to do so until
the pulsar distance can be measured more accurately.  We expect that
the Shapiro gravitational propagation delay will yield additional
tests of relativistic gravity within a few more years.

\acknowledgements{The three authors gratefully acknowledge financial support from the
  US National Science Foundation.  Arecibo Observatory is operated by
  Cornell University under cooperative agreement with the NSF. We
  thank Joseph Swiggum for assistance with analyses of glitches in the
  pulsar population; and C . M. Ewers\footnote{Deceased}, A. de la Fuente, J.T. Green, 
  and Z. Pei for assistance with observations.}

\newpage
\begin{deluxetable}{lcccccc}
\tablecolumns{7}
\tablecaption{Parameters of Observing Systems}
\tablehead{
\colhead{Name}  &  \colhead{Dates} & \colhead{Total}             & \colhead{Frequency}  & \colhead{Time} & \colhead{TOA} & \colhead{Number of}  \\
\colhead{}             & \colhead{}                                &  \colhead{Bandwidth} &  \colhead{Channels}   &  \colhead{Resolution}     &  \colhead{Uncertainty}     &  \colhead{TOAs}    \\
 \colhead{} &  \colhead{} &  \colhead{(MHz)} &  \colhead{} &  \colhead{$(\mu$s)}  & \colhead{$(\mu$s)} \\
}
\startdata
Mark I\dotfill    & 1981 -- 1984 & 16 &  64 & 125 & 20 & 1719 \\
Mark II\dotfill  & 1985 -- 1989 &  8 &  32 & 125 & 31 & 1015  \\
Mark III\dots & 1988 -- 2003 & 40 & 32 & 640 & 16 &  2473  \\
WAPP\tablenotemark{a}\dotfill      & 2003 -- 2006 & 100  & 512 & 58 & 14  & 2443 \\ 
\enddata 
\tablenotetext{a}{Three or four WAPPS were simultaneously employed in nonoverlapping 
frequency bands. The listed numbers refer to a single WAPP.}
\label{table:obsys}
\end{deluxetable}

\newpage

\begin{deluxetable}{lll}
\tablecolumns{3}
\tablecaption{Astrometric and Spin Parameters}
\tablehead{

\colhead{Parameter}  &   \colhead{Value\tablenotemark{a}} \\

}
\startdata
$t_0$ (MJD)\tablenotemark{b}\dotfill   & 52984.0 \\
$\alpha$ (J2000)\dotfill & $19^h 15^m 27\fs 99928(9)$ & \\
$\delta$ (J2000)\dotfill &   16\arcdeg 06\arcmin 27\farcs3871(13) & \\
$\mu_{\alpha}$ (mas yr$^{-1}$)\dots &   $-$1.43(13) &  \\
$\mu_{\delta}$ (mas yr$^{-1}$)\dotfill  &   $-$0.70(13)  & \\
$f$ (s$^{-1}$)\dotfill  & 16.94053778563(15)  & \\
$\dot{f}$ (s$^{-2}$)\dotfill  & $-$2.4761(9) $\times 10^{-15}$ &  \\
Glitch epoch (MJD)\dotfill & 52770(20) \\
$\Delta f$ (s$^{-1}$)\dotfill & 6.2(2) $\times 10^{-10}$ \\

\enddata 
\tablenotetext{a}{Figures in parentheses represent estimated
  uncertainties in the last quoted digit.  The estimated uncertainties range
  from $(3-10)\times$ the formal fitted uncertainties, in order to also reflect
  variations resulting from different assumptions regarding timing noise, etc.}
  \tablenotetext{b}{This quantity is the epoch of the next six measurements
  tabulated here.}
 \label{table:astromfits}
\end{deluxetable}

\begin{deluxetable}{ll}
\tablecolumns{2}
\tablecaption{Orbital Parameters}
\tablehead{

\colhead{Parameter}  &   \colhead{Value\tablenotemark{a}}  \\

}
\startdata
$T_0 $ (MJD)\dotfill & 52144.90097841(4)   \\
$x\equiv a_1 \sin i$  (s)\dots & 2.341782(3)  \\
$e$ \dotfill &   0.6171334(5)  \\
$P_b$ (d)\dotfill  &   0.322997448911(4)   \\
$\omega_0$ (deg)\dotfill   &  292.54472(6)   \\
$\langle\dot{\omega}\rangle$ (deg / yr)\dotfill & 4.226598(5)   \\
$\gamma$ (ms) \dotfill & 4.2992(8)   \\
$\dot{P}_b $\dotfill  & $-$2.423(1) $\times 10^{-12}$   \\

\enddata 
\tablenotetext{a}{Figures in parentheses represent estimated
  uncertainties in the last quoted digit. The estimated uncertainties range
  from $(2-6)\times$ the formal fitted uncertainties, in order to also reflect
  variations resulting from different assumptions regarding timing noise, etc.}
 \label{table:orbfits}
\end{deluxetable}

\newpage
\begin{figure}
\includegraphics[trim=0.2in 0.7in  0 0.7in ,clip,scale=0.7]{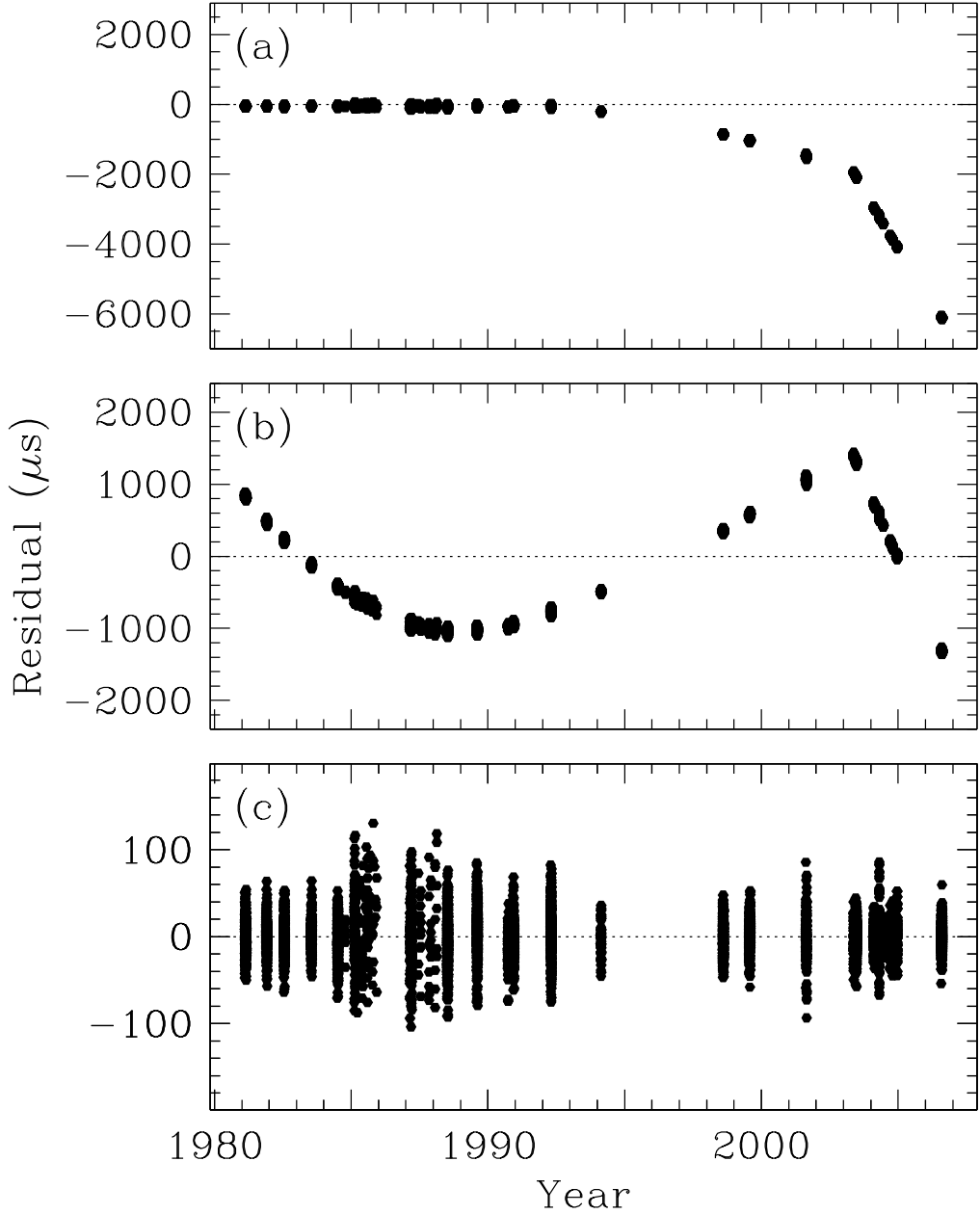}
\caption{Timing residuals for PSR B1913+16.  (a) Residuals from a fit for data 
before mid-1992. The  glitch in May  2003 can be recognized by a distinct change in 
slope of the residuals versus time.  The apparent change in mid-1992 is 
much smaller and may or may not involve a discrete event.  (b) Residuals 
from a fit of all data, holding astrometric and orbital parameters fixed at the 
values in Tables \ref{table:astromfits} and \ref{table:orbfits}; fitting for pulsar
 frequency and spin-down rate, $f$ and $\dot{f}$; and not allowing for higher 
 order frequency derivatives or glitches.  The glitch in May 2003 is evident as a 
 sharp discontinuity.  (c) Residuals from the full timing fit, including higher order 
 frequency derivatives and the glitch.}
\label{fig:glitch}
\end{figure}

\newpage


\begin{thebibliography}{}

\bibitem[Arzoumanian et al.(1994)]{aet94} Arzoumanian, Z., 
Nice, D.~J., Taylor, J.~H., \& Thorsett, S.~E.\ 1994, \apj, 422, 671 

\bibitem[Barker \& O'Connell(1975a)]{boa75} Barker, B.~M., \& O'Connell, R.~F.\ 1975a, \prd, 12, 329 

\bibitem[Barker \& Oconnell(1975b)]{bob75} Barker, B.~M., \& O'Connell, R.~F.\ 1975b, \apjl, 199, L25 

\bibitem[Chukwude 
\& Urama(2010)]{cu10} Chukwude, A.~E., \& Urama, J.~O.\ 2010, \mnras, 910, in press

\bibitem[Clifton \& Weisberg(2008)]{cw08} Clifton, T., \& Weisberg, J.~M.\ 2008, \apj, 679, 687 

\bibitem[Cognard  \& Backer(2004)]{cb04} Cognard, I., \& Backer, D.~C.\ 2004,
 \apjl, 612, L125 
 
 \bibitem[Cordes \& Downs(1985)]{cd85} Cordes, J.~M., \& Downs, G.~S.\ 1985, 
 \apjs, 59, 343 

\bibitem[Damour \& Ruffini(1974)]{dr74} Damour, T., \& Ruffini, R.\ 1974, Academie des Sciences Paris 
Comptes Rendus Serie Sciences Mathematiques, 279, 971 

\bibitem[Damour \& Deruelle(1986)]{dd86}Damour, T.~\&  Deruelle, N.\ 1986, Ann. Inst. H. 
Poincare (Phys. Theorique), 44, 263

\bibitem[Damour \& Taylor(1991)]{dt91} Damour, T., \& Taylor, J.~H.\ 1991, \apj, 366, 501 

\bibitem[Damour \& Taylor(1991)]{dt92} Damour, T., \& Taylor, J.~H.\ 1992, Phys. Rev D, 45, 1840

\bibitem[Damour(2009)]{d07} Damour, T.\ 2009, in  	{\it{Physics of Relativistic Objects in Compact Binaries: 
From Birth to Coalescence}}, eds. M. Colpi, U. Moschella, P. Casella, A. Possenti, \& V. Gorini;
Springer Netherlands. Astrophysics and Space Science Library, 359, 1;  
preprint available at arXiv:0707.0749 [gr-qc] 

\bibitem[Edwards et al.(2006)]{eet06} Edwards, R.~T., Hobbs, 
G.~B., \& Manchester, R.~N.\ 2006, \mnras, 372, 1549 

\bibitem[Esposito-Farese(2009)]{esp09} Esposito-Farese, G.\ 
2009, arXiv:0905.2575 [gr-qc]

\bibitem[Ghez et al.(2008)]{ghez08} Ghez, A.~M., et al.\ 2008, 
\apj, 689, 1044 

\bibitem[Gillessen et al.(2009)]{get09} Gillessen, S., 
Eisenhauer, F., Trippe, S., Alexander, T., Genzel, R., Martins, F., 
\& Ott, T.\ 2009, \apj, 692, 1075 

\bibitem[Hobbs et al.(2004)]{het04} Hobbs, G., Lyne, A.~G., 
Kramer, M., Martin, C.~E., \& Jordan, C.\ 2004, \mnras, 353, 1311 

\bibitem[Hobbs et al.(2006)]{het06} Hobbs, G.~B., Edwards, 
R.~T., \& Manchester, R.~N.\ 2006, \mnras, 369, 655 

\bibitem[Hulse  \& Taylor(1975)]{ht75} Hulse, R.~A., \& Taylor, J.~H.\ 1975, \apjl, 195, L51 

\bibitem[Janssen et al.(2008)]{jet08} Janssen, G.~H., Stappers, B.~W., 
Kramer, M., Nice, D.~J., Jessner, A., Cognard, I., \& Purver, M.~B.\ 2008, \aap, 490, 753 

\bibitem[Kramer(1998)]{k98} Kramer, M.\ 1998, \apj, 509, 856 



\bibitem[Melatos et al.(2008)]{mel08} Melatos, A., Peralta, 
C., \& Wyithe, J.~S.~B.\ 2008, \apj, 672, 1103

\bibitem[Peters and Mathews (1963)]{pm63} Peters. P. C., and 
Mathews, J.\ 1963, Phys Rev, 131, 435

\bibitem[Reid et al.(2009)]{ret09} Reid, M.~J., et al.\ 2009, 
\apj, 700, 137 

\bibitem[Sch{\"o}nrich et al.(2010)]{set10} Sch{\"o}nrich, 
R., Binney, J., \& Dehnen, W.\ 2010, \mnras, 403, 1829 


\bibitem[Shapiro (1964)]{sha64} Shapiro, I.~I. \ 1964, \prl, 13, 789

\bibitem[Stairs et al.(2006)]{set06} Stairs, I.~H., Thorsett, 
S.~E., Dewey, R.~J., Kramer, M., \& McPhee, C.~A.\ 2006, \mnras, 373, L50 

\bibitem[Taylor et al.(1976)]{tet76} Taylor, J.~H., Hulse, 
R.~A., Fowler, L.~A., Gullahorn, G.~E., \& Rankin, J.~M.\ 1976, \apjl, 206, L53 

\bibitem[Taylor et al.(1979)]{tet79} Taylor, J.~H., Fowler, L.~A., \& McCulloch, P.~M.\ 1979, 
\nat, 277, 437 

\bibitem[Taylor \& Weisberg(1982)]{tw82} Taylor, J.~H., \& Weisberg, J.~M.\ 1982, \apj, 253, 908 

\bibitem[Taylor \& Weisberg(1989)]{tw89} Taylor, J.~H., \& Weisberg, J.~M.\ 1989, \apj, 345, 434 

\bibitem[Taylor et al.(1992)]{tet92} Taylor, J.~H., 
Wolszczan, A., Damour, T., \& Weisberg, J.~M.\ 1992, \nat, 355, 132 

\bibitem[Thorsett et al.(2005)]{tet05} Thorsett, S.~E., 
Dewey, R.~J., \& Stairs, I.~H.\ 2005, \apj, 619, 1036 

\bibitem[Urama et al.(2006)]{ulw06} Urama, J.~O., Link, B., 
\& Weisberg, J.~M.\ 2006, \mnras, 370, L76 

\bibitem[Weisberg \& Taylor(1981)]{wt81} Weisberg, J.~M., \& Taylor, J.~H.\ 1981, General 
Relativity and Gravitation, 13, 1 

\bibitem[Weisberg et al.(1989)]{wet89} Weisberg, J.~M., Romani, R.~W., \& Taylor, J.~H.\ 1989, 
\apj, 347, 1030 

\bibitem[Weisberg \& Taylor(2002)]{wt02} Weisberg, J.~M., \& Taylor, J.~H.\ 2002, \apj, 576, 942 

\bibitem[Weisberg et al.(2008)]{wet08} Weisberg, J.~M., 
Stanimirovi{\'c}, S., Xilouris, K., Hedden, A., de la Fuente, A., Anderson, 
S.~B., \& Jenet, F.~A.\ 2008, \apj, 674, 286 

\end{thebibliography}
\end{document}